\documentclass{article}

\usepackage[centertags]{amsmath}
\usepackage{amssymb,amsfonts,eucal,oldgerm,textcomp}
\usepackage[text={6in,8.25in},centering]{geometry}
\usepackage[colorlinks=true,citecolor=red,linkcolor=blue,urlcolor=blue,pdftex]{hyperref}
\usepackage{chicago}

\newcommand{\skipline}[1][1]{\vspace*{#1\baselineskip}}
\newcommand{\nth}[1]{#1$^{\text{th}}$} 

\newcommand{\half}{\frac{1}{2}}

\newtheorem{theorem}{Theorem}[subsection]

\newtheorem{defn}[theorem]{Definition}
\newtheorem{prop}[theorem]{Proposition}
\newtheorem{lemma}[theorem]{Lemma}

\title{A Simple Proof of the Uniqueness of the Einstein Field Equation
  in All Dimensions\thanks{This paper was submitted to \emph{Physics
      Review Letters}, Jan 2016.  I thank Robert Geroch for many
    enjoyable conversations on these matters, during the course of
    which many of the ideas in this paper were germinated and in some
    cases brought to full fruition.}}

\author{Erik Curiel\thanks{\textbf{Author's address}: Munich Center
    for Mathematical Philosophy, Ludwig-Maximilians-Universit\"at,
    Ludwigstra{\ss}e 31, 80539 M\"unchen, Deutschland;
    \textbf{email}: \href{mailto:erik@strangebeautiful.com}
    {\texttt{erik@strangebeautiful.com}}}}

\begin{document}
\thispagestyle{empty}
\maketitle

\skipline

\begin{quote}
  \begin{center}
    \textbf{ABSTRACT}
  \end{center}  

  The standard argument for the uniqueness of the Einstein field
  equation is based on Lovelock's Theorem, the relevant statement of
  which is restricted to four dimensions.  I prove a theorem similar
  to Lovelock's, with a physically modified assumption: that the
  geometric object representing curvature in the Einstein field
  equation ought to have the physical dimension of stress-energy.  The
  theorem is stronger than Lovelock's in two ways: it holds in all
  dimensions, and so supports a generalized argument for uniqueness;
  it does not assume that the desired tensor depends on the metric
  only up second-order partial-derivatives, that condition being a
  consequence of the proof.  This has consequences for understanding
  the nature of the cosmological constant and theories of
  higher-dimensional gravity.  Another consequence of the theorem is
  that it makes precise the sense in which there can be no
  gravitational stress-energy tensor in general relativity.  Along the
  way, I prove a result of some interest about the second jet-bundle
  of the bundle of metrics over a manifold.
\end{quote}

\skipline

The Einstein field equation, $G_{ab} = 8 \pi \gamma T_{ab}$ (where
$\gamma$ is Newton's gravitational constant) consists of an object
representing the curvature of spacetime (the Einstein tensor,
$G_{ab}$) equated with the stress-energy tensor of material fields
($T_{ab}$).  The standard proof of the uniqueness of the equation
invokes the classic theorem by \citeN{lovelock72},
\begin{theorem}
  \label{thm:lovelock}
  Let $(\mathcal{M},\; g_{ab})$ be a four-dimensional spacetime.  In a
  coordinate neighborhood of a point $p \in \mathcal{M}$, let
  $\Theta_{\alpha \beta}$ be the components of a tensor concomitant of
  $\{g_{\lambda \mu} ; \; g_{\lambda \mu,\nu} ; \; g_{\lambda \mu ,
    \nu \rho} \}$ such that
    \[
    \nabla^n \Theta_{nb} = 0.
    \]
    Then
    \[
    \Theta_{ab} = r G_{ab} + q g_{ab},
    \]
    where $\nabla_a$ is the derivative operator associated with the
    metric $g_{ab}$, $G_{ab}$ is the Einstein tensor, and $q$ and $r$
    are constants.
\end{theorem}
The restriction to four dimensions is essential for the result.  In
higher dimensions, there are other tensors satisfying the theorem.
(Those tensors are not linear in the second-order partial-derivatives
of the metric as the Einstein tensor is.)  Those tensors form the
basis of so-called Lovelock gravity theories
\cite{lovelock71,padmanabhan-kothawala-lanczos-lovelock-mods}.

In this note, I sketch the proof of the following:
\begin{theorem}
  \label{thm:only-einstein-tens}
  The only two covariant-index, divergence-free, concomitants of the
  metric that are homogeneous of weight zero are constant multiples of
  the Einstein tensor.
\end{theorem}
There is a subtle but important difference between Lovelock's original
theorem and my result, one with interesting consequences.  Lovelock
did not require the concomitant to be homogeneous, the assumption
capturing the idea that the desired concomitant has the physical
dimension of stress-energy (as I explain below).  The theorem is thus
weaker than Lovelock's in one sense.  It also, however, makes it
stronger in two important senses: the assumption of being second-order
in the metric is not required, but follows from the proof; and perhaps
more importantly, my result does not depend on the dimension of the
manifold, proving uniqueness of the Einstein field equation in all
dimensions, not just four.

First, I lay down the needed definitions.  (From hereon, I use the
Geroch-Newman-Penrose abstract-index notation; see \citeNP{wald-gr}.)
\begin{defn}
  \label{def:concom}
  For two fiber bundles $(\mathcal{B}_1, \, \mathcal{M}, \, \pi_1)$
  and $(\mathcal{B}_2, \, \mathcal{M}, \, \pi_2)$ over the same base
  space $\mathcal{M}$, a mapping
  $\chi : \mathcal{B}_1 \rightarrow \mathcal{B}_2$ is a
  \emph{concomitant} if
  \[
  \chi (\phi^*_1 (u_1)) = \phi^*_2 (\chi (u_1))
  \]
  for all $u_1 \in \mathcal{B}_1$ and all diffeomorphisms $\phi$ from
  $\mathcal{M}$ to itself, where $\phi^*_i$ is the natural
  diffeomorphism induced by $\phi$ on the bundle space
  $\mathcal{B}_i$.
\end{defn}
This definition can be generalized to take account of how a
concomitant can depend on differentials of the fiber bundle
$\mathcal{B}$ that is its domain, based on the $\nth$-order jet bundle of
$\mathcal{B}$, $J^n \mathcal{B}$.  There is a natural projection
$\theta^{n,m} : J^n \mathcal{B} \rightarrow J^m \mathcal{B}$ (for
$0 < m < n)$, characterized by taking the Taylor expansion that
defines the $n$-jet and ``dropping all terms above order $m$''.
\begin{defn}
  \label{defn:nth-concoms}
  An \emph{$n^{\text{th}}$-order concomitant} ($n$ a strictly positive
  integer) from $\mathcal{B}_1$ to $\mathcal{B}_2$ (bundles over the
  same base space $\mathcal{M})$ is a smooth mapping
  $\chi : J^n \mathcal{B}_1 \rightarrow \mathcal{B}_2$ such that for
  all $u \in J^n \mathcal{B}_1$ and diffeomorphisms $\phi$ from
  $\mathcal{M}$ to itself
  \begin{enumerate}
      \item $\phi^*_2 (\chi(u)) = \chi (\phi^*_n (u))$
      \item there is no $(n-1)^{\text{th}}$-order concomitant
    $\chi' : J^{n-1} \mathcal{B}_1 \rightarrow \mathcal{B}_2$
    satisfying $\chi (u) = \chi' (\theta^{n,n-1} (u))$ for all
    $u \in J^n \mathcal{B}_1$
  \end{enumerate}
\end{defn}
A zeroth-order concomitant (or just `concomitant' for short, when no
confusion will arise) is one satisfying definition~\ref{def:concom}.
An important property of concomitants is that, in a limited sense,
they are transitive.
\begin{prop}
  \label{prop:con-composition}  
  If $\chi_1: J^n \mathcal{B}_1 \rightarrow \mathcal{B}_2$ is an
  $n^{\text{th}}$-order concomitant and $\chi_2: \mathcal{B}_2
  \rightarrow \mathcal{B}_3$ is a smooth mapping, where
  $\mathcal{B}_1$, $\mathcal{B}_2$ and $\mathcal{B}_3$ are geometric
  bundles over the same base space, then $\chi_2 \circ \chi_1$ is an
  $n^{\text{th}}$-order concomitant if and only if $\chi_2$ is a
  zeroth-order concomitant.
\end{prop}
This follows immediately from the definition of \nth{n}-order
concomitants and the properties of the natural lifts of
diffeomorphisms from a base space to a jet bundle.  Finally, a
concomitant is \emph{homogeneous of weight $w$} if for any constant
scalar field $\xi$
\[
\chi (\phi^*_1 (\xi u)) = \xi^w \phi^*_2 (\chi (u))
\]
This definition makes sense, as we consider only bundles of linear and
affine objects in this paper.

We now explicate the structure of the first two jet bundles of the
bundle of metrics over a manifold.  Two metrics $g_{ab}$ and $h_{ab}$
are in the same 1-jet at a point if and only if they have the same
associated covariant derivative operator at that point.  To see this,
first note that, if they are in the same 1-jet, then
$\hat{\nabla}_a (g_{bc} - h_{bc}) = 0$ at that point for all
derivative operators.  Thus, for the derivative operator $\nabla_a$
associated with, say, $g_{ab}$, $\nabla_a (g_{bc} - h_{bc}) = 0$, but
$\nabla_a g_{bc} = 0$, so $\nabla_a h_{bc} = 0$ at that point as well.
Similarly, if the two metrics are equal and share the same associated
derivative operator $\nabla_a$ at a point, then
$\hat{\nabla}_a (g_{bc} - h_{bc}) = 0$ at that point for all
derivative operators, since their difference will be identically
annihilated by $\nabla_a$, and $g_{ab} = h_{ab}$ at the point by
assumption.  Thus they are in the same 1-jet.  This proves that all
and only geometrically relevant information contained in the 1-jets of
Lorentz metrics on $\mathcal{M}$ is encoded in the fiber bundle over
spacetime the values of the fibers of which are ordered pairs
consisting of a metric and the metric's associated derivative operator
at a spacetime point.

The second jet bundle over $\mathcal{B}_{\text{\small g}}$ has a
similarly interesting structure.  Clearly, if two metrics are in the
same 2-jet, then they have the same Riemann tensor at the point
associated with the 2-jet, since the result of doubly applying to it
an arbitrary derivative operator (not the Levi-Civita one associated
with the metric) at the point yields the same tensor.  Assume now that
two metrics are in the same 1-jet and have the same Riemann tensor at
the associated spacetime point.  If it follows that they are in the
same 2-jet, then essentially all and only geometrically relevant
information contained in the 2-jets of Lorentz metrics on
$\mathcal{M}$ is encoded in the fiber bundle over spacetime the points
of the fibers of which are ordered triplets consisting of a metric,
the metric's associated derivative operator and the metric's Riemann
tensor at a spacetime point.  To demonstrate this, it suffices to show
that if two Levi-Civita connections agree on their respective Riemann
tensors at a point, then the two associated derivative operators are
in the same 1-jet of the bundle whose base-space is $\mathcal{M}$ and
whose fibers consist of the affine spaces of derivative operators at
the points of $\mathcal{M}$ (because they will then agree on the
result of application of themselves to their difference tensor, and
thus will be in the 2-jet of the same metric at that point).

Assume that, at a point $p$ of spacetime, $g_{ab} = \tilde{g}_{ab}$,
$\nabla_a = \tilde{\nabla}_a$ (the respective derivative operators),
and $R^a {}_{bcd} = \tilde{R}^a {}_{bcd}$ (the respective Riemann
tensors).  Let $C^a {}_{bc}$ be the symmetric difference-tensor
between $\nabla_a$ and $\tilde{\nabla}_a$, which is itself 0 at $p$ by
assumption.  Then by definition $\nabla_{[b} \nabla_{c]} \xi^a = R^a
{}_{bcn} \xi^n$ for any vector $\xi^a$, and so at $p$
\begin{equation*}
  \begin{split}
    R^c {}_{abn} \xi^n &= \nabla_{[a} \tilde{\nabla}_{b]} \xi^c \\
    &= \nabla_a (\nabla_b \xi^c + C^c {}_{bn} \xi^n) -
    \tilde{\nabla}_b  \nabla_a\xi^c \\
    &= \nabla_a \nabla_b \xi^c + \nabla_a (C^c {}_{bn} \xi^n) -
    \nabla_b \nabla_a \xi^c - C^c {}_{bn} \nabla_a \xi^n + C^n {}_{ba}
    \nabla_n\xi^c
  \end{split}
\end{equation*}
but $\nabla_b \nabla_c \xi^a - \nabla_c \nabla_b \xi^a = 2 R^a
{}_{bcn} \xi^n$ and $C^a {}_{bc} = 0$, so expanding the only remaining
term gives
\[
\xi^n \nabla_a C^c {}_{bn} = 0
\]
for arbitrary $\xi^a$ and thus $\nabla_a C^b {}_{cd} = 0$ at $p$; by
the analogous computation, $\tilde{\nabla}_a C^b {}_{cd} = 0$ as well.
It follows immediately that $\nabla_a$ and $\tilde{\nabla}_a$ are in
the same 1-jet over $p$ of the affine bundle of derivative operators
over $\mathcal{M}$.  We have proven
\begin{theorem}
  \label{thm:1-2-jet-metric}
  $J^1 \mathcal{B}_{\text{\small g}}$ is naturally diffeomorphic to
  the fiber bundle over $\mathcal{M}$ whose fibers consist of pairs
  $(g_{ab}, \, \nabla_a)$, where $g_{ab}$ is the value of a Lorentz
  metric field at a point of $\mathcal{M}$, and $\nabla_a$ is the
  value of the covariant derivative operator associated with $g_{ab}$
  at that point.  $J^2 \mathcal{B}_{\text{\small g}}$ is naturally
  diffeomorphic to the fiber bundle over $\mathcal{M}$ whose fibers
  consist of triplets $(g_{ab}, \, \nabla_a, \, R_{abc} {}^d)$, where
  $g_{ab}$ is the value of a Lorentz metric field at a point of
  $\mathcal{M}$, and $\nabla_a$ and $R_{abc}{}^d$ are respectively the
  covariant derivative operator and the Riemann tensor associated with
  $g_{ab}$ at that point.
\end{theorem}
It follows immediately that there is a first-order concomitant from
$\mathcal{B}_{\text{\small g}}$ to the geometric bundle
$(\mathcal{B}_\nabla, \, \mathcal{M}, \, \pi_\nabla,$ $\iota_\nabla)$
of derivative operators, \emph{viz}., the mapping that takes each
Lorentz metric to its associated derivative operator.  Likewise, there
is a second-order concomitant from $\mathcal{B}_{\text{\small g}}$ to
the geometric bundle
$(\mathcal{B}_{\text{\small Riem}}, \, \mathcal{M}, \,
\pi_{\text{\small Riem}},$
$\iota_{\text{\small Riem}})$ of tensors with the same index structure
and symmetries as the Riemann tensor, \emph{viz}., the mapping that
takes each Lorentz metric to its associated Riemann tensor.  (This is
the precise sense in which the Riemann tensor associated with a given
Lorentz metric is ``a function of the metric and its partial
derivatives up to second order''.)  It is easy to see, moreover, that
both concomitants are homogeneous of degree 0.

It follows from theorem~\ref{thm:1-2-jet-metric} and
proposition~\ref{prop:con-composition} that a concomitant of the
metric will be second order if and only if it is a zeroth-order
concomitant of the Riemann tensor:
\begin{prop}
  \label{prop:0-concoms-riemann}
  A concomitant of the metric is second-order if and only if it can be
  expressed as a sum of terms consisting of constants multiplied by
  the Riemann tensor, the Ricci tensor, the Gaussian scalar curvature,
  and contractions and products of these with the metric itself.
\end{prop} 

Now, in order to make precise the idea of having the physical
dimension of stress-energy, recall that in general relativity all the
fundamental units one uses to define stress-energy, namely time,
length and mass, can themselves be defined using only the unit of
time; these are so-called geometrized units.  This guarantees that
units of mass and length scale in precisely the same manner as the
time-unit when new units of time are chosen by multiplying the
time-unit by some fixed real number $\lambda^{-\frac{1}{2}}$.  (The
reason for the inverse square-root will become clear in a moment).
Thus, a duration of $t$ time-units would become
$t\lambda^{-\frac{1}{2}}$ of the new units; an interval of $d$ units
of length would likewise become $d\lambda^{-\frac{1}{2}}$ in the new
units, and $m$ units of mass would become $m\lambda^{-\frac{1}{2}}$ of
the new units.  This justifies treating all three of these units as
``the same'', and so expressing acceleration, say, in inverse
time-units.  To multiply the length of all timelike vectors
representing an interval of time by $\lambda^{-\frac{1}{2}}$, however,
is equivalent to multiplying the metric by $\lambda$ (and so the
inverse metric by $\lambda^{-1}$), and indeed such a multiplication is
the standard way one represents a change of units in general
relativity.  This makes physical sense as the way to capture the idea
of physical dimension: all physical units, the ones composing the
dimension of any physical quantity, are geometrized in general
relativity in the most natural formulation, and so depend only on the
scale of the metric itself.  By Weyl's theorem, however, a metric
times a constant represents exactly the same physical phenomena as the
original metric \cite[ch.~2, \S1]{malament-fnds-gr-ngt}.

Now, the proper dimension of a stress-energy tensor can be determined
by the demand that the Einstein field-equation,
$G_{ab} = 8 \pi \gamma T_{ab}$, remain satisfied when one rescales the
metric by a constant factor.  $\gamma$ has dimension
$\frac{\mbox{(length)}^3} {\mbox{(mass)(time)}^2}$, and so in
geometrized units does not change under a constant rescaling of the
metric.  Thus $T_{ab}$ ought to transform exactly as $G_{ab}$ under a
constant rescaling of the metric.  A simple calculation shows that
$G_{ab} \space (= R_{ab} - \half R g_{ab})$ remains unchanged under
such a rescaling.  Thus, a necessary condition for a tensor to
represent stress-energy is that it remain unchanged under a constant
rescaling of the metric.  It follows that the concomitant at issue
must be homogeneous of weight 0 in the metric, whatever order it may
be.

We must still determine the order of the required concomitant.  In
fact, the weight of a homogeneous concomitant of the metric suffices
to fix the differential order of that concomitant.\footnote{I thank
  Robert Geroch for pointing this out to me.}  This can be seen as
follows, as exemplified by the case of a two covariant-index,
homogeneous concomitant $S_{ab}$ of the metric.  A simple calculation
based on definition~\ref{defn:nth-concoms} and on the fact that the
concomitant must be homogeneous shows that the value at a point
$p \in \mathcal{M}$ of an $n^{\text{th}}$-order concomitant $S_{ab}$
can be written in the general form
\begin{equation}
  \label{eq:Sab-form}
  S_{ab} = \sum_\alpha k_\alpha \, g^{qx} \ldots g^{xr} \left(
    \widetilde{\nabla}_x^{(n_1)} g_{qx} \right) \ldots \left(
    \widetilde{\nabla}_x^{(n_i)} g_{xr} \right)
\end{equation}
where: $\widetilde{\nabla}_a$ is any derivative operator at $p$
\emph{other} than the one naturally associated with $g_{ab}$; `$x$' is
a dummy abstract index; `$\widetilde{\nabla}_x^{(n_i)}$' stands for
$n_i$ iterations of that derivative operator (obviously each with a
different abstract index); $\alpha$ takes its values in the set of all
permutations of all sets of positive integers $\{ n_1, \ldots, n_i \}$
that sum to $n$, so $i$ can range in value from 1 to $n$; the
exponents of the derivative operators in each summand themselves take
their values from $\alpha$, \emph{i}.\emph{e}., they are such that
$n_1 + \cdots + n_i = n$ (which makes it an $n^{\text{th}}$-order
concomitant); for each $\alpha$, $k_\alpha$ is a constant; and there
are just enough of the inverse metrics in each summand to contract all
the covariant indices but $a$ and $b$.

Now, a combinatorial calculation shows
\begin{prop}
  \label{prop:nth-concom-factor}
  If, for $n \geq 2$, $S_{ab}$ is an $n^{\text{th}}$-order homogeneous
  concomitant of $g_{ab}$, then to rescale the metric by the constant
  real number $\lambda$ multiplies $S_{ab}$ by $\lambda^{n - 2}$.
\end{prop}
In other words, the only such homogeneous $n^{\text{th}}$-order
concomitants must be of weight $\lambda - 2$.\footnote{The exponent
  $(n - 2)$ in this result depends crucially on the fact that $S_{ab}$
  has only two indices, both covariant.  One can generalize the result
  for tensor concomitants of the metric of any index structure.  A
  slight variation of the argument, moreover, shows that there does
  not in general exist a homogeneous concomitant of a given
  differential order from a tensor of a given index structure to one
  of another structure---one may not be able to get the number and
  type of the indices right by contraction and tensor multiplication
  alone.}  So if one knew that $S_{ab}$ were multiplied by, say,
$\lambda^4$ when the metric was rescaled by $\lambda$, one would know
that it had to be a sixth-order concomitant.  In particular, $S_{ab}$
does not rescale when $g_{ab} \rightarrow \lambda g_{ab}$ only if it
is a second-order homogeneous concomitant of $g_{ab}$,
\emph{i}.\emph{e}., (by theorem~\ref{thm:1-2-jet-metric} and
proposition~\ref{prop:0-concoms-riemann}) a zeroth-order concomitant
of the Riemann tensor.  There follows from
proposition~\ref{prop:con-composition}
\begin{lemma}
  \label{lem:riem-0th-concom-0-homog}
  A 2-covariant index concomitant of the Riemann tensor is homogeneous
  of weight zero if and only if it is a zeroth-order concomitant.
\end{lemma}
Thus, such a tensor has the physical dimension of stress-energy if and
only if it is a zeroth-order concomitant of the Riemann tensor.  It is
striking how powerful the physically motivated assumption that the
required object have the physical dimensions of stress-energy: it
guarantees that the required object will be a second-order concomitant
of the metric.

Now, it follows from proposition~\ref{prop:0-concoms-riemann} that the
only possibilities for geometrical objects to place on the lefthand
side of a field equation that would play the role of the Einstein
field equation are linear combinations of the Ricci tensor and the
scalar curvature multiplied by the metric.  The only covariantly
divergence-free, linear combinations of those two quantities, however,
are constant multiples of the Einstein tensor $G_{ab}$.  (To see this,
note that if there were another, say $k_1 R_{ab} + k_2 R g_{ab}$ for
constants $k_1$ and $k_2$, then
$k_1 R_{ab} + k_2 R g_{ab} - 2k_2 G_{ab}$ would also be divergence
free, but that expression is just a constant multiple of the Ricci
tensor.)  This proves theorem~\ref{thm:only-einstein-tens}.  A benefit
of the proof is that it gives real geometrical and physical insight
into the result, insight not provided by Lovelock's original proof of
theorem~\ref{thm:lovelock}, which consists of several pages of
unilluminating coordinate-based, brute-force calculation.

Theorem~\ref{thm:only-einstein-tens} shows the uniqueness of the
Einstein field equation in all dimensions.  The theorem is similar to
Lovelock's result, but different in four important ways. The first
difference is that I require the concomitant of the metric to be
homogeneous of weight zero.  The physical interpretation of this is
that the desired tensor has the physical dimensions of stress-energy,
as is the case for the Einstein tensor, and as must be the case for
any tensor that one would equate to a material stress-energy tensor to
formulate a field equation.  This provides a physical interpretation
to the conditions of the theorem that Lovelock's theorem lacks.  It
also leads to the second difference: one does not need to assume that
the desired concomitant is second-order; that property falls naturally
out of the proof.

The third difference is that the theorem holds in all dimensions, not
just in four.  It follows that, in dimensions other than four, the
so-called Lovelock tensors ar not homogeneous of weight zero, and so
do not have the physical dimension of stress-energy.  Thus, if one
wants to construct a field equation that equates a linear combination
of them to the stress-energy tensor of ordinary matter, as Lovelock
theories of gravity do, then the coupling constants cannot be
dimensionless like Newton's gravitational constant; the physical
dimension of each coupling constant will be determined by the physical
dimension of the Lovelock tensor it multiplies.

The fourth difference is that the addition of constant multiples of
the metric is not allowed. I interpret that to mean that any
cosmological-constant term must be construed as part of the total
stress-energy tensor of spacetime, and so, in particular, the
cosmological constant itself must have the physical dimensions of
(mass)$^2$.

Finally, theorem~\ref{thm:only-einstein-tens} has another natural
interpretation: it shows in a precise and rigorous sense the
nonexistence of a gravitational stress-energy tensor.  If there were
such a thing, we would expect it to depend on curvature, and so be
zero in and only in flat spacetimes.  Constant multiples of the
Einstein tensor, however, are not appropriate candidates for the
representation of gravitational stress-energy: the Einstein tensor
will be zero in a spacetime having a vanishing Ricci tensor but a
non-trivial Weyl tensor; such spacetimes, however, can manifest
phenomena, \emph{e}.\emph{g}., pure gravitational radiation in the
absence of ponderable matter, that one naturally wants to say possess
gravitational energy in some (necessarily non-localized) form or
other.\footnote{This argument, by the way, obviates the criticism of
  the claim that gravitational stress-energy ought to depend on the
  curvature.  The critics point out that that would make gravitational
  stress-energy depend on second-order partial derivatives of the
  field potential, whereas all other known forms of stress-energy
  depend only on terms quadratic in the first partial derivatives of
  the field potential.  It is, however, exactly second-order,
  homogeneous concomitants of the metric that possess terms quadratic
  in the first partials.  The rule is that the order of a homogeneous
  concomitant is the sum of the exponents of the derivative operators
  when the concomitant is represented in the form of
  equation~\eqref{eq:Sab-form}.}  (See
\citeNP{curiel-geom-objs-nonexist-sab-uniq-efe} for extended
discussion of this and other intpretational issues raised in this
paper.)

\bibliographystyle{chicago}

\end{document}